%% file: paper.tex
\documentclass[letterpaper,twocolumn,screen,10pt]{article}
\usepackage{usenix}

\input{includes/packages}

\input{includes/settings}

\if\showcomments1
    \usepackage[colorinlistoftodos,textsize=tiny]{todonotes}
\else
    \usepackage[disable]{todonotes}
\fi

\begin{document}

\title{\titleinfo}

\if\shownames1
    \author{\authorinfo \\ \institutioninfo}
\else
    \author{\submissioninfo}
\fi

\if\showcomments1
	\onecolumn
    \setcounter{page}{0}
    \listoftodos{}
    \clearpage
    \twocolumn
    \setcounter{page}{1}
\fi

\if\showpagenumbers0
	\pagestyle{empty}
\fi

\maketitle

\input{sections/abstract}

\input{sections/introduction}
\input{sections/background}
\input{sections/design}
\input{sections/implementation}
\input{sections/evaluations}

\input{sections/limitations}
\input{sections/related-work}
\input{sections/conclusion}

\label{lastpage}

{\small
\bibliographystyle{plain}
\bibliography{bibs/paper}}

\label{totalpage}

\end{document}

%% file: includes/packages.tex
\usepackage{xspace} 
\usepackage{enumitem}
\usepackage{color}
\usepackage{colortbl}
\usepackage{pifont}

\usepackage{epstopdf}
\usepackage{graphicx}
\usepackage[labelfont=bf, textfont=bf]{caption}
\usepackage{subcaption}
\usepackage{pgfplots}
\usepackage{pgfplotstable}

\usepackage{siunitx}
\usepackage{ifthen}
\usepackage{algorithm}
\usepackage[noend]{algpseudocode}

\usepackage{listings}
\usepackage[utf8]{inputenc}

\usepackage{booktabs}  
\usepackage{tabu}
\usepackage{array}
\usepackage{multirow}
\usepackage{diagbox}
\usepackage{tabularx}
\usepackage{amssymb}
\usepackage[dvipsnames]{xcolor}
\usepackage{rotating}
\usepackage[normalem]{ulem}

\usepackage{balance}
\usepackage{microtype}

\usepackage[english]{babel}
\usepackage{blindtext}

\usepackage[most]{tcolorbox}

\usepackage[capitalise]{cleveref}

\usepackage{xurl}

%% file: includes/settings.tex
\def\setuppreprint{1}
\def\setupcameready{0}

\newcommand{\name}{\textsc{OptiNIC}\xspace}

\newcommand{\titleinfo}{\name{}: A Resilient and Tail-Optimal RDMA NIC for Distributed ML Workloads} 

\newcommand{\authorinfo}{\em Ertza Warraich, Ali Imran$^{\dagger}$, Annus Zulfiqar$^{\dagger}$, Shay Vargaftik$^{*}$, Sonia Fahmy, Muhammad Shahbaz$^{\dagger}$}

\newcommand{\institutioninfo}{
Purdue University~~~$^{*}$Broadcom~~~$^{\dagger}$University of Michigan\vspace{-10pt}
}
\newcommand{\submissioninfo}{\em Submission \#69 \\ {\normalsize \pageref{lastpage} Pages Body, \pageref{totalpage} Pages Total}}

\if\setupcameready1
	\def\shownames{1} 										
	\def\showpagenumbers{0}                                 
    \def\showcomments{0} 									
\else
	\def\shownames{1}										
	\def\showpagenumbers{1}                                 
    \if\setuppreprint1										
        \def\showcomments{0}
    \else
        \def\showcomments{1}
    \fi
\fi

\newcommand{\ie}{i.e.}
\newcommand{\eg}{e.g.}

\setlist[itemize]{topsep=4pt, itemsep=4pt, parsep=1.5pt}

\newcommand{\cmark}{\ding{51}}%
\newcommand{\xmark}{\ding{55}}%
\useunder{\uline}{\ul}{}

%% file: sections/abstract.tex
\begin{abstract}
As distributed machine learning (ML) workloads scale to thousands of GPUs connected by high-speed interconnects, tail latency in collective communication has become a major bottleneck.
Existing RDMA transports, such as RoCE, IRN, SRNIC, and Falcon, enforce strict reliability and in-order delivery, relying on retransmissions and packet sequencing to ensure correctness.
While these approaches work well for general-purpose workloads, they introduce complexity and latency that scale poorly in ML, where even rare packet delays can stall entire model pipelines.

We present \name{}, a domain-specific RDMA transport that revisits traditional reliability guarantees based on ML's tolerance for partial or missing data.
\name{} eliminates retransmissions and in-order delivery from the NIC, enabling a best-effort, out-of-order transport model for RDMA.
Unlike traditional RDMA, which signals completion only after complete data delivery, \name{} introduces adaptive timeouts to trigger forward progress when data may be lost or delayed.
\name{} retains standard congestion control mechanisms (\eg, DCQCN, EQDS, or Swift) while shifting loss recovery to the ML pipeline itself (\eg, via the Hadamard Transform and Erasure Coding).

Our evaluation shows that \name{} improves time-to-accuracy (TTA) by 2$\times$ and increases throughput by 1.6$\times$ for training and inference, respectively, across two public clouds (\ie, Hyperstack and CloudLab). 
\name{} also lowers 99th-percentile latency by 3.5$\times$, cuts BRAM usage by 2.7$\times$, and nearly doubles NIC resilience to faults---delivering a resilient, tail-optimized RDMA transport purpose-built for distributed ML workloads.
\end{abstract}

%% file: sections/introduction.tex
\section{Introduction}
\label{sec:introduction}

As distributed machine learning (ML) workloads scale across thousands of GPUs connected by high-speed 100--400G fabrics, the performance bottleneck has shifted decisively from compute to communication.~\cite{shah2023taccl,li2024thc}
Collective operations (such as AllReduce, AllGather, and All-to-All) have become critical synchronization points in both data-parallel and model-parallel training and inference pipelines~\cite{warraich2025optireduce,wang2024towards,zhou2025extensible,uec2025spec}.
These operations demand tight coordination among workers, where even minor tail delays in the communication fabric can stall overall progress~\cite{sapio2021scaling}. 
As a result, {\em tail latency}, not average throughput, has emerged as the dominant barrier to scaling ML workloads efficiently across large clusters~\cite{warraich2025optireduce,wang2024towards}.

To address this communication bottleneck, the community has introduced a range of optimizations. 
Systems like NCCL~\cite{jeaugey2017nccl}, RCCL~\cite{rccl}, and MSCCL~\cite{msccl} apply both algorithmic~\cite{shah2023taccl,shah2025msccl++,venkata2024unified} and hardware-aware~\cite{graham2016scalable,lao2021atp,sapio2021scaling} techniques to accelerate collectives. 
At the same time, compression methods such as gradient sparsification~\cite{wangni2017gradient,fei2021efficient,renggli2019sparcml} and quantization~\cite{alistarh2017qsgd} reduce bandwidth demands by exploiting the statistical nature of stochastic gradient descent (SGD)~\cite{fei2021efficient,alistarh2017qsgd}.
These approaches are built on a key observation: {\em ML workloads are statistically robust}. 
They tolerate approximation, noise, and even bounded data loss without compromising final model accuracy~\cite{dettmers2022gpt3, frantar2022gptq, xiao2023smoothquant}. 

Despite these advances,
the underlying transport layer has largely remained general-purpose and overly conservative~\cite{mittal2018revisiting,wang2023srnic,singhvi2025falcon}.
In modern ML clusters, RDMA is the dominant communication substrate, typically implemented via RoCE or its derivatives~\cite{guo2016rdma,gangidi2024rdma}.
These transports enforce strict reliability and in-order delivery, relying on retransmissions, packet sequencing, and lossless flow control mechanisms like Priority Flow Control (PFC) to ensure correctness. 
While appropriate for traditional distributed systems (like key-value stores or databases)~\cite{dragojevic2014farm,wei2020xstore,qiao2023hermit,raghavan2023cornflakes}, these mechanisms are increasingly sub-optimal for ML. 
Their reliance on complete delivery as a precondition for forward progress introduces latency-critical paths that do not scale.
A single packet loss can cascade into Go-Back-N retransmission storms 
or PFC-induced head-of-line blocking, stalling pipelines across the entire cluster~\cite{gangidi2024rdma,wang2023srnic}.

In response, several recent systems have begun rethinking RDMA transport. 
IRN~\cite{mittal2018revisiting} removes PFC by enabling in-NIC loss recovery through selective repeat, bitmap tracking, and SACK-based retransmissions. 
While this design improves cluster scalability, it inflates per-QP state and adds reordering complexity in the NIC. 
SRNIC~\cite{wang2023srnic} simplifies the NIC datapath by removing \textsf{WQE} caching and onloading retransmissions and reordering to host software, improving QP density and reducing NIC memory pressure. 
UCCL~\cite{zhou2025extensible} pushes this idea further by onloading the entire transport control plane---including congestion control, flow scheduling, and multipath routing---into software, treating the NIC as a streamlined datapath.

In contrast, Falcon~\cite{singhvi2025falcon} takes the opposite approach: it embraces NIC complexity by integrating fast retransmissions, delay-based congestion control, and multipath routing directly into hardware. 
While Falcon performs well under loss, it increases the NIC state and vulnerability to hardware faults. 
At the same time, the Ultra Ethernet Consortium (UEC) proposes a clean-slate design for AI workloads~\cite{uec2025spec,hoefler2025ultra}, introducing features like packet spraying, hybrid congestion control, and fast loss detection. 
However, like Falcon and IRN-based approaches, UEC's proposed transport preserves strict reliability semantics---requiring full delivery before forward progress.

The above-mentioned systems represent important steps forward, but they all retain a common assumption: that packet loss is rare and must be recovered before computation can continue~\cite{mittal2018revisiting,wang2023srnic,singhvi2025falcon,zhou2025extensible}.
They preserve the long-standing semantic that {\em forward progress is gated on complete delivery}. 
At the ML scale, however, this assumption no longer holds. 
What seems like rare loss at a single node becomes frequent across thousands of workers synchronizing in parallel~\cite{warraich2025optireduce,wang2024towards}.
These losses accumulate at collective barriers, where even a single straggler can stall the entire operation. 
This is a classic case of ``tail at scale''~\cite{dean2013tail}, worsened by transport-layer mechanisms that insist on full recovery before making progress.

In this paper, we ask: {\em If ML workloads can tolerate partial loss and reordering, why enforce strict delivery guarantees at the NIC at all?} 
If the application is already robust to bounded loss, why not remove the mechanisms that wait for full delivery entirely?

We present \name{}, a domain-specific RDMA transport that rethinks reliability and forward progress from the ground up. 
\name{} eliminates retransmissions and in-order delivery from the NIC, forwarding best-effort, out-of-order packets directly to application memory. 
Crucially, \name{} replaces delivery-based progress with a new primitive: {\em adaptive timeouts}. 
Rather than waiting for every packet to arrive, the receiver proceeds once a fixed time elapses---even if some data is missing. 
These timeouts are coordinated across peers and tuned to the collective's structure (\eg, Ring, Tree, BCube), providing consistent, time-bounded semantics for progress in lossy networks.

\name{} preserves compatibility with existing RDMA infrastructure. 
It retains standard congestion control mechanisms (such as DCQCN~\cite{dcqcn}, EQDS~\cite{olteanu2022edge}, or Swift~\cite{kumar2020swift}) and maintains IB verbs semantics~\cite{nvidia-rdma-manual,infiniband-manual}, while keeping the RDMA programming model intact. 
Rather than recovering from packet loss within the transport, \name{} bounds its impact and shifts recovery to the ML stack, where lightweight redundancy mechanisms (such as the Hadamard Transform~\cite{warraich2025optireduce}) can reconstruct missing data efficiently. 
Timeout tuning, credit/window management,
and error handling remain in software, preserving flexibility without adding NIC complexity.

This architectural shift significantly simplifies the NIC. 
\name{} eliminates reorder buffers, retransmission queues, and per-packet sequencing logic, cutting NIC BRAM usage by 2.7$\times$ and nearly doubling mean-time-between-failure (MTBF) by removing fault-prone state. 
It also prevents tail latencies caused by recovery delays at the cluster scale.

We further evaluate \name{} across a range of ML workloads and public cloud environments (\S\ref{sec:evaluation}).
In clusters running on Hyperstack and CloudLab, \name{} delivers 1.8--2.5$\times$ speedups for collectives across message sizes and topologies. 
For end-to-end training, \name{} improves time-to-accuracy (TTA) by 2$\times$ using ZeRO-3 parallelism, boosts inference throughput by 1.6$\times$, and reduces time-to-first-token (TTFT) tail latency---a key measure of LLM responsiveness---by 3.5$\times$, all while preserving model accuracy.

In short, \name{} challenges the long-held assumption that reliable delivery is a necessary precondition for correctness in distributed ML systems. 
By replacing delivery-based progress with timeout-driven semantics tailored to ML, \name{} enables a new class of transport designs: simple, stateless, and resilient---co-designed for the unique demands of modern machine learning.

%% file: sections/background.tex
\begin{table*}[t]
\centering
\footnotesize
\resizebox{0.999\textwidth}{!}{
\begin{tabular}{@{}l|l|l|l|l|l|>{\columncolor[HTML]{cfe2f3}}l}
\toprule
\textbf{Feature} 
    & \textbf{RoCE} 
    & \textbf{IRN}~\cite{mittal2018revisiting} 
    & \textbf{SRNIC}~\cite{wang2023srnic} 
    & \textbf{Falcon}~\cite{singhvi2025falcon}
    & \textbf{UCCL}~\cite{zhou2025extensible} 
    & \textbf{\name{}} \\ 
\midrule
Transport Reliability        
    & Go-Back-N (HW)        
    & Selective Repeat (HW) 
    & Selective Repeat (SW) 
    & Selective Repeat (HW) 
    & Selective Repeat (SW) 
    & Best Effort \\
Packet Reordering            
    & No/Dropped
    & Buffered in NIC       
    & Software Reordering   
    & Buffered in NIC 
    & Software Reordering 
    & Offset Based \\
Congestion Control           
    & Hardware              
    & Hardware              
    & Hardware    
    & Hardware 
    & Software              
    & Hardware \\
Priority Flow Control           
    & Required              
    & Not Required          
    & Not Required          
    & Not Required 
    & Not Required 
    & Not Required \\
Target Workloads             
    & General RDMA          
    & General RDMA          
    & RDMA + ML       
    & RDMA + ML + HPC 
    & ML Collectives 
    & ML Collectives \\
\midrule
\textbf{Key Focus}          
    & High performance      
    & +Network efficiency   
    & +Connection scalability 
    & +Programmable CC  
    & +Programmable transport 
    & +Tail optimality \\
\bottomrule
\end{tabular}
}
\vspace{-8pt}
\caption{\bf Evolution of RDMA transport designs: from reliability-centric to tail-optimal. \name{} eliminates retransmission and reordering machinery, and uses best-effort, offset-based placement to support large-scale ML collectives.}
\label{tab:rdma-evolution}
\vspace{-14pt}
\end{table*}

\section{Background and Motivation}
\label{sec:background}

\subsection{Communication Bottlenecks in ML Workloads}
\label{ssec:comm-bottlenecks}

Distributed ML workloads rely on fine-grained, structured communication between GPUs to synchronize computation across a cluster. 
These communication patterns are dominated by collectives such as AllToAll (\textsf{AA}), AllReduce (\textsf{AR}), AllGather (\textsf{AG}), and ReduceScatter (\textsf{RS}), which are invoked on every iteration of training or every decoding step during inference.
The specific collective topology and frequency are determined by the parallelism strategy employed---data, model, pipeline, tensor, or hybrid~\cite{vllm-paper,zhao2023pytorch}.

In data parallelism, models are replicated across workers and gradients are synchronized using \textsf{AR}. 
More advanced variants, such as Fully Sharded Data Parallelism (FSDP)~\cite{zhao2023pytorch} or ZeRO-3~\cite{rajbhandari2020zero}, reduce memory usage by partitioning the model state, but introduce additional collectives, \textsf{AG} and \textsf{RS}, within each training step.
In model or pipeline parallelism, activations are passed between layer partitions across devices, resulting in fine-grained, latency-sensitive exchanges. 
Tensor parallelism introduces intra-layer collectives: for example, outputs of split matrix multiplications must be gathered across GPUs to proceed. 
Inference workloads further introduce complications with cache lookups, sequence-level slicing, and context merging, frequently relying on \textsf{AA} or \textsf{AG} collectives.

\begin{figure}[t]
  \centering
  \includegraphics[width=\linewidth]{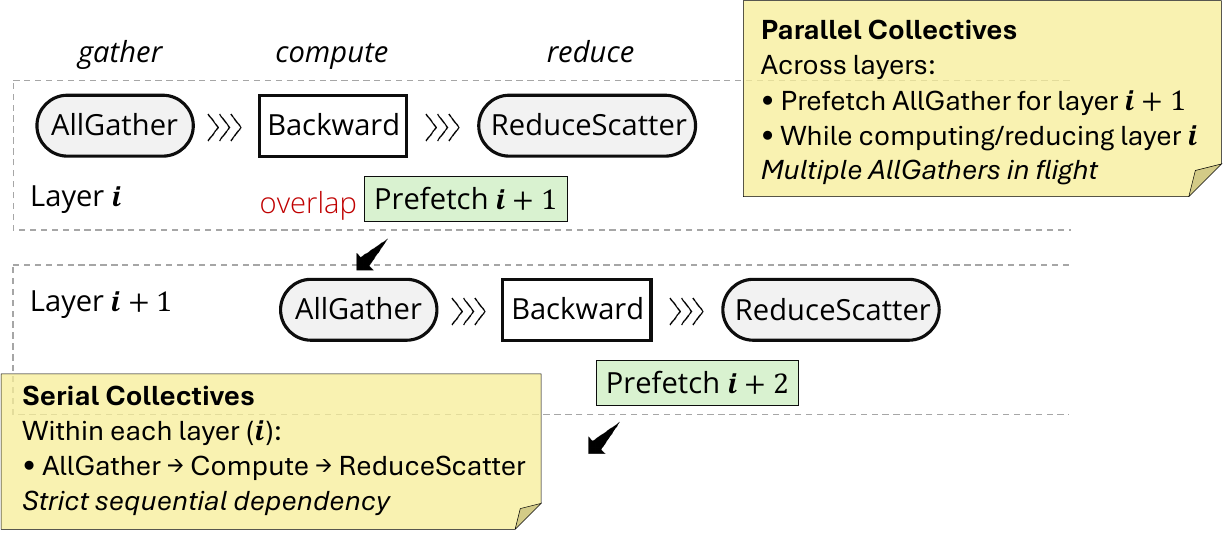}
  \vspace{-18pt}
    \caption{Overlapping intra-layer and inter-layer collective patterns during FSDP backward pass.}
    \vspace{-15pt}
  \label{fig:collective-comm}
\end{figure}

Most often, workloads combine these strategies. 
\Cref{fig:collective-comm} illustrates the backward pass of an FSDP pipeline, where intra-layer collectives (\textsf{AG} $\rightarrow$ compute $\rightarrow$ \textsf{RS}) are chained together, and inter-layer prefetches run concurrently. 
This mixture of serial and parallel collectives creates intricate synchronization dependencies that define the critical path.

As models scale to hundreds of billions of parameters~\cite{li2024thc,warraich2025optireduce} and clusters scale to thousands of GPUs~\cite{gangidi2024rdma}, these collectives increasingly dominate end-to-end performance. 
Even a single delayed packet in one GPU's \textsf{AG} can stall the entire iteration. 
Studies show that collectives can account for 50--70\% of total runtime in such systems~\cite{warraich2025optireduce,sapio2021scaling}.
While bandwidth requirements are well-understood, the bottleneck is not throughput but tail latency---specifically, the delay incurred by the slowest GPU in each synchronization round.

This effect is especially pronounced during inference. 
In multi-query batching or sequence-parallel pipelines, collectives are triggered at sub-millisecond granularity, with little opportunity to amortize communication delays. 
Time-to-first-token (TTFT) is directly impacted by per-step stalls, even if only a few packets are delayed. 
Moreover, the data exchanged in these collectives---intermediate tensors, activation fragments, KV cache blocks---is often small, redundant, or transient~\cite{yang2022inflless, fu2024serverlessllm}.
Yet today's transports still enforce strict delivery semantics on every packet, waiting for full and in-order delivery before triggering progress. 
These semantics ignore ML's tolerance to loss (\S\ref{ssec:ml-loss}) and impose tail-latency penalties disproportionate to the impact of missing data.

\begin{figure}[t]
  \centering
  \begin{subfigure}[b]{0.39\linewidth}
    \centering
    \includegraphics[width=\textwidth]{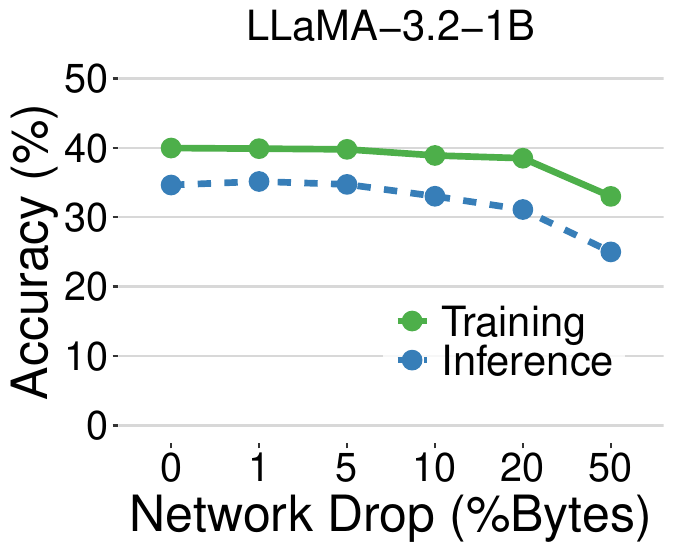}
    \vspace{-17pt}
    \caption{\normalfont Training and inference on the ARC dataset.}
    \label{fig:training}
  \end{subfigure}
  \hfill
  \begin{subfigure}[b]{0.59\linewidth}
    \centering
    \includegraphics[width=\textwidth]{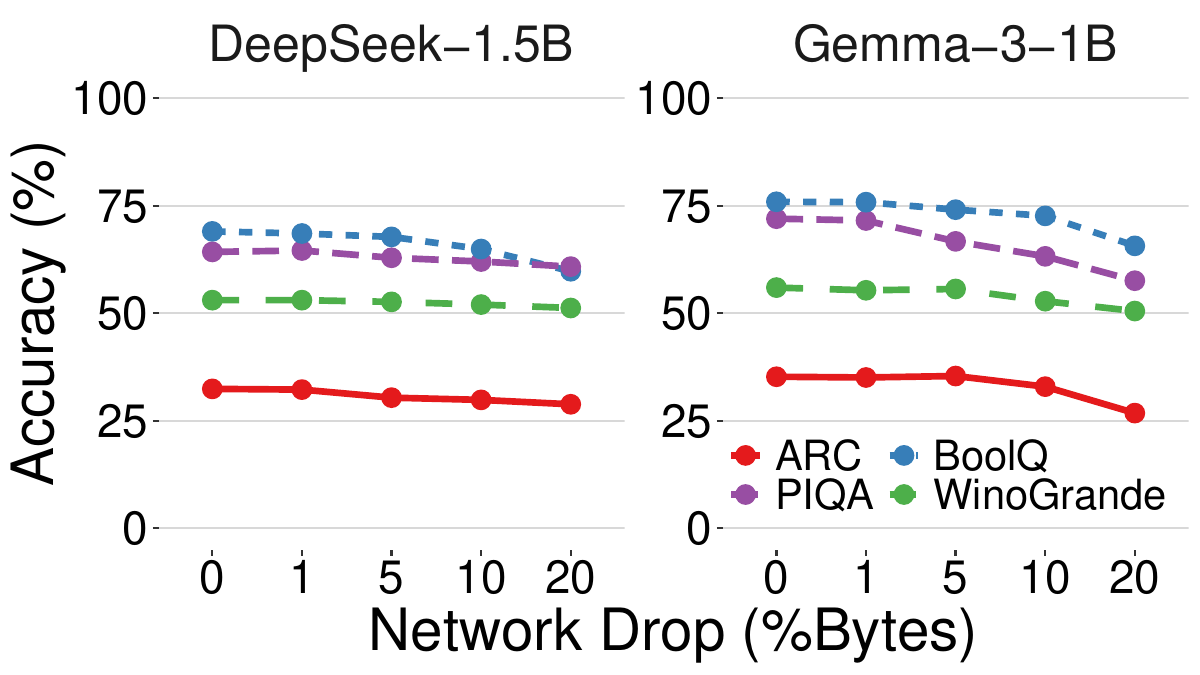}
	\vspace{-17pt}
    \caption{\normalfont Inference-only accuracy using multiple datasets (ARC, BoolQ, and more).}
    \label{fig:inference}
  \end{subfigure}
  \vspace{-8pt}
  \caption{\bf Training and inference accuracy of all models remains stable under partial network drops ($\leq$ 5\%).}
  \label{fig:loss-resilience}
  \vspace{-18pt}
\end{figure}

\subsection{ML Workloads are Resilient to Loss}
\label{ssec:ml-loss}

The communication bottleneck in ML arises not only from message frequency but also from a mismatch between transport semantics and application needs. 
ML workloads are not fragile distributed systems. 
Rather, they are designed to be robust to approximation, randomness, and partial data~\cite{li2024thc,wang2024towards}.

Stochastic gradient descent (SGD) inherently absorbs noise across iterations~\cite{fei2021efficient}.
Prior systems have leveraged this property to reduce bandwidth through quantization~\cite{alistarh2017qsgd}, gradient sparsification~\cite{fei2021efficient}, or reduced-precision formats like \textsf{bfloat16}~\cite{fei2021efficient, alistarh2017qsgd}. 
In-network aggregation frameworks such as SHARP perform approximate or lossy reductions directly in the dataplane, showing that full-precision delivery is not necessary for convergence~\cite{graham2016scalable}.

This resilience extends beyond gradients. 
Activations, attention maps, and routing metadata are often recomputed or subsampled in subsequent steps. 
In MoE models~\cite{shazeer2017outrageously,rajbhandari2022deepspeed},
missing expert outputs may be ignored or replaced via fallback paths. 
In self-attention~\cite{vaswani2017attention},
partial loss in key/value tensors may have a negligible impact due to the smoothing behavior of softmax layers. 
\Cref{fig:loss-resilience} shows that across a variety of large-language models (LLMs) and datasets, both training and inference accuracy remain stable even at 5\% drop rates.

Even when loss occurs, it is not uniformly damaging. Many tensors are padded, sparse, or partially redundant. From an application perspective, ML pipelines do not require every packet to arrive; rather, they require only enough data to complete the current step. This suggests a different progress model: one that favors bounded, timely delivery over strict reliability. While best-effort delivery may introduce nondeterminism, such variability is already present in large-scale training pipelines, and techniques such as per-step logging or structured redundancy
can aid debugging and reproducibility.

\subsection{The Cost of Reliable (RDMA) Transports}
\label{ssec:cost-of-reliability}

Despite loss tolerance of ML workloads, the transport layer remains conservative. 
RDMA transports like RoCE enforce strict reliability by default: Go-Back-N retransmissions, in-order delivery, and PFC to avoid loss~\cite{mittal2018revisiting,wang2023srnic}. 
These mechanisms are implemented entirely in hardware. 
NICs maintain per-connection state, including retry counters, sequence numbers, window logic, and reorder buffers. 
They use acknowledgment packets to detect loss, timers to trigger recovery, and congestion windows to pace the sender. 
Even the widely-used NVIDIA NCCL stack uses RDMA's reliable RC queue pairs (QPs) with control/data QPs per peer, tightly coupling transport reliability with application progress~\cite{jeaugey2017nccl}.

Reliable semantics work well for key-value stores or RPCs, but they scale poorly for ML~\cite{warraich2025optireduce,wang2024towards}.
A single lost packet can block an entire collective.
Worse, reliability mechanisms turn rare events into protocol-level delays. 
PFC-induced backpressure causes head-of-line blocking. 
Reordering logic inflates NIC memory usage. Retransmissions inject traffic bursts.
In large ML jobs, where collectives synchronize thousands of workers, such events are no longer rare---they are expected.
What appears as a one-in-a-thousand loss at a single node becomes one per step across the cluster.

Recent designs have tried to reduce this complexity. 
IRN replaces Go-Back-N with selective repeat, using bitmaps and selective ACKs to recover lost packets more efficiently~\cite{mittal2018revisiting}.
SRNIC offloads reordering and retransmissions to software and eliminates the \textsf{WQE} cache, reducing NIC state~\cite{wang2023srnic}. 
UCCL moves the entire transport control plane into software, using the NIC purely as a datapath~\cite{zhou2025extensible}.
Falcon enhances the NIC instead, tightly integrating loss recovery, congestion control, and multipath routing for tail performance under stress~\cite{singhvi2025falcon}.

These efforts vary in architecture but share a core assumption: that loss must be detected and corrected before progress. 
\Cref{tab:rdma-evolution} compares these designs. 
All still enforce strict delivery semantics and treat forward progress as a function of full data arrival. 
This model is fundamentally misaligned with ML workloads, where approximate or delayed recovery is not only acceptable---it is preferable to waiting.

\subsection{Reliability Hurts Fault Tolerance}
Finally, enforcing reliability reduces the fault tolerance of the NIC. 
Transport-layer mechanisms---retry logic, sequence tracking, congestion windows---are stored in NIC SRAM, tightly coupled with datapath execution. 
These stateful elements are vulnerable to soft errors, transient failures, and silent corruption~\cite{amd-seu-reports,keller2021terrestrial}. 
At a cluster scale, even conservative mean-time-between-failure (MTBF) estimates lead to frequent failures. 
A stuck timer, corrupted sequence number, or missed completion can stall a QP indefinitely, halting collectives and triggering global backoff.

This fragility is unnecessary. 
ML workloads can make forward progress without complete delivery. 
Rather than hardening unreliable machinery inside the NIC, we eliminate it. 
\name{} discards retransmissions, in-order enforcement, and per-packet tracking altogether. 
Instead, it forwards best-effort packets directly to memory and uses a timeout mechanism to signal progress when no new packets arrive.
Recovery, if needed, is handled in software via structured redundancy (\eg, Hadamard Transform~\cite{warraich2025optireduce}), \S\ref{ssec:recovery-mitigation}.

Our architecture reduces per-QP state to just 20 bytes: no retry counters, timers, reorder buffers, or flow windows. 
Only minimal congestion control metadata remains. 
As we show in (\S\ref{ssec:microbenchmarks}), this design reduces NIC BRAM usage by 2.7$\times$, nearly doubles MTBF in hardware fault models, and eliminates the tail stalls caused by reliability logic. 

%% file: sections/design.tex
\section{Design of \name{}}
\label{sec:design}

We present \name{}, a resilient, tail-optimal RDMA architecture for ML that eliminates retransmissions and in-order delivery at the NIC. 
Instead of tying progress to reliable delivery, \name{} introduces bounded completion semantics: each operation completes within an application-specified timeout, and the NIC signals partial completion to enable timely progress---even when some data is lost. 

\name{} occupies a new point in the RDMA transport design space (\Cref{tab:qp-comparison})---dropping reliability and ordering guarantees while retaining connection state, hardware packetization, and congestion control. 
These choices reflect its goal: to minimize tail latency and protocol overhead while preserving the RDMA programming model.

\begin{table}[t]
\centering
\footnotesize
\renewcommand{\arraystretch}{1.1}
\begin{tabular}{cccccc}
\toprule
\multirow{2}{*}{\textbf{QP Type}} & \multicolumn{1}{c}{\textbf{Offloaded}} & \multicolumn{1}{c}{\textbf{Reli-}} & \multicolumn{1}{c}{\textbf{In-Order}} & \multirow{2}{*}{\textbf{CC}} \\
& \textbf{Packetization} & \textbf{ability} & \textbf{Delivery} & \\
\midrule
\textsf{RC} & \cmark & \cmark & \cmark & \cmark \\
\textsf{UC} & \cmark & \xmark & \cmark & \xmark \\
\textsf{UD} & \xmark & \xmark & \xmark & \xmark \\
\rowcolor[HTML]{cfe2f3}
\textbf{\name{}: \textsf{XP}} & \cmark & \xmark & \xmark & \cmark \\
\bottomrule
\end{tabular}
\vspace{-5pt}
\caption{Comparison of RDMA QP types~\cite{nvidia-rdma-manual} and \name{} along key transport features. 
RC ensures reliability and ordering but incurs high tail latency; UC drops reliability but lacks congestion control (CC); UD offers no hardware support.
\name{}'s XP (eXpress Path) fills the gap: it drops reliability and ordering while retaining connection state, offloaded packetization, and CC.}
\vspace{-16pt}
\label{tab:qp-comparison}
\end{table}

We begin by describing the transport architecture of \name{} (\S\ref{ssec:transport-arch}), covering its design for data delivery, bounded completion semantics, and congestion control.
Next, we describe a software-level recovery mechanism (\S\ref{ssec:recovery-mitigation}), which enables model correctness even when packets are lost or partially delivered.
We then present two deployment mappings (\S\ref{ssec:mapping-nics}), showing how \name{} can be realized with minimal changes to existing RDMA NICs: (1) by modifying SRNIC~\cite{wang2023srnic} to remove reliability machinery, and (2) using off-the-shelf RoCE NICs with the UC transport.

\subsection{\name{}'s Transport Architecture for RDMA}
\label{ssec:transport-arch}
\name{} reimagines the standard RDMA transport abstractions---data delivery, completion, and congestion control---to support a lossy, time-bounded execution model tailored for ML workloads. 
While these services exist in all RDMA transports~\cite{nvidia-rdma-manual}, their implementation in \name{} is explicitly designed to tolerate packet loss, avoid reordering overheads, and guarantee forward progress within application-defined time bounds. 
Each transport component is simplified to operate without retransmissions or ordering guarantees, while preserving the core RDMA programming model.

In \name{}, senders issue standard RDMA operations, which are fragmented into self-describing packets that can be delivered and placed independently of arrival order (\S\ref{sssec:data-delivery}). 
Completion is decoupled from reliable delivery: each work request entry (\textsf{WQE}) includes a timeout, and the NIC signals completion either when all fragments are received, or the timer expires (\S\ref{sssec:bounded-compl}). 
This bounded approach enables timely progress reporting and downstream recovery. 
Congestion control remains intact: \name{} supports existing ECN-, delay-, or credit-based logics by separating rate regulation from reliability, allowing pacing to operate cleanly over a best-effort substrate (\S\ref{sssec:cc}).

\begin{tcolorbox}[
    colback=blue!5!white,
    colframe=black,
    title={},
    fonttitle=\bfseries,
    fontupper=\footnotesize, 
    sharp corners,
    boxrule=0.4mm,
    coltitle=black,
    enhanced jigsaw,
    drop shadow={black!50!white}
]
\textbf{INFO: Key RDMA Concepts~\cite{nvidia-rdma-manual}.}
\begin{itemize}[leftmargin=*]
	\item RDMA transports use Work Queue Entries (\textsf{WQEs}) to describe operations such as \textsf{SEND}, \textsf{RECV}, or \textsf{WRITE}. Each \textsf{WQE} corresponds to a message and resides on a queue pair (\textsf{QP}). When an operation completes, the NIC posts a Completion Queue Entry (\textsf{CQE}) to notify the application.
	\item One-sided verbs like \textsf{WRITE} allow a sender to directly place data into the receiver's memory without coordination. Two-sided verbs like \textsf{SEND}/\textsf{RECV} require both sender and receiver to post matching \textsf{WQEs}.
	\item Packets within a message are usually sequenced using Packet Sequence Numbers (\textsf{PSNs}). In \name{}, we instead use a per-message sequence number called \textsf{wqe\_seq} to identify operations and support timeout and preemption logic. Remote memory addresses are communicated via the \textsf{RETH} (RDMA Extended Transport Header) in one-sided operations.
\end{itemize}
\end{tcolorbox}

\subsubsection{Data Delivery Semantics.~}
\label{sssec:data-delivery}
The core function of RDMA transport is to move data from source to destination memory via direct memory access (DMA). 
For large messages, the NIC splits data into MTU-sized packets that must be placed correctly in the destination buffer. 
Traditional RDMA transports---such as Reliable Connected (RC) and Unreliable Connected (UC), \Cref{tab:qp-comparison}---rely on strict in-order delivery for correct placement: only the first packet carries the full remote address and offset, while later packets infer their position using implicit packet sequencing. 
This ordering assumption forces reliable transports to either buffer out-of-order packets (\eg, with selective repeat) or retransmit them (\eg, with Go-Back-N), introducing latency, memory pressure, and tail amplification under loss.

In \name{}, we remove these dependencies by treating out-of-order arrival as the common case. 
We eliminate reordering logic entirely and instead ensure correct placement through {self-describing packets}, each of which carries sufficient metadata to be placed independently of other packets.

\paragraph{\em Self-Describing Packets.} To place a packet correctly, the receiver must know two things: (1) which message the packet belongs to, and (2) where within the target buffer the payload should be written. 
Existing transports do not provide this information per packet. 
For one-sided verbs (like RDMA WRITE), only the first packet includes a \textsf{RETH} header with the virtual address, remote key (rkey), and total length; subsequent packets rely on in-order arrival to infer their offset. 
Two-sided verbs like \textsf{SEND}/\textsf{RECV} behave similarly: the receiver has the base address from its posted \textsf{WQE},
but packets carry no explicit offset---again relying on strict ordering. 

This model breaks under loss or reordering. 
RC must buffer until gaps are filled; UC simply drops out-of-order packets. 
Neither can safely place packets that arrive out of order.
\name{} addresses this by making every packet self-describing. 
Each fragment carries metadata needed for direct placement:

\begin{itemize}[leftmargin=*]
	\item For one-sided operations, each packet includes the full \textsf{RETH} header: the virtual address (with offset) and rkey.\vspace{-3pt}
	\item For two-sided \textsf{SEND}/\textsf{RECV}, each packet includes a byte offset into the pre-posted receive buffer.
\end{itemize}

This allows the receiver to perform in-place DMA on arrival---without buffering, reordering, or inferring offset from PSN.
The NIC simply extracts the offset from the packet header and writes the payload to memory. 
This design supports correct placement regardless of arrival order and applies uniformly to one-sided and two-sided operations.

\paragraph{\em Out-of-Order Delivery Across Messages.} 
Traditional RDMA semantics also require that messages be delivered in order: packets from a new message cannot be processed until the previous message completes. 
This works under reliable delivery, where missing fragments are eventually retransmitted. 
In \name{}, however, packets may be lost permanently. 
Waiting for missing fragments would stall the QP indefinitely.

To allow forward progress without buffering, \name{} introduces a single-active-message model. 
Each packet carries a \textsf{wqe\_seq} identifier to indicate which message it belongs to. 
The receiver maintains a single expected \textsf{wqe\_seq} and processes packets as follows:

\begin{itemize}[leftmargin=*]
	\item If the packet's \textsf{wqe\_seq} matches the expected value, it is part of the active message and is placed immediately.\vspace{-3pt}
	\item If the packet's \textsf{wqe\_seq} is greater, the sender has moved on. \vspace{-3pt}
	The receiver finalizes the previous message and begins processing the new one.\vspace{-3pt}
	\item If the \textsf{wqe\_seq} is less, the packet belongs to a completed (or timed out) message and is dropped.
\end{itemize}

This keeps receiver state bounded: only one active message is tracked per QP, and no per-message buffering is required. 
The arrival of a new message acts as an implicit timeout for the previous one, allowing the receiver to progress earlier.

\paragraph{\em Late Packet Handling.} Once a message is completed---either by receiving its final fragment or by timeout---the NIC advances the expected \textsf{wqe\_seq}, clears associated state, and posts a \textsf{CQE}.
Any packets that arrive afterward with the old sequence number are immediately dropped. 
This ensures correctness even in the presence of delayed fragments or multipath reordering: late packets cannot corrupt application memory or confuse the completion logic. 

\subsubsection{(Bounded) Completion Semantics.~}
\label{sssec:bounded-compl}
Traditional RDMA transports define completion based on reliable delivery: an operation completes when all its fragments are received and acknowledged. 
This model assumes eventual delivery and ties progress to in-order arrival and retransmissions. 
In \name{}, however, these assumptions no longer hold. 
There are no retransmissions, and packets may be lost permanently. 
To ensure forward progress without reliability, \name{} introduces a new model: {\em bounded completion semantics}, where each operation completes within a timeout and reports partial progress if necessary.

\name{} preserves the notion of completion familiar to RDMA developers.
On the sender side, a \textsf{WQE} is marked complete once all fragments have been transmitted---no acknowledgments are required.
On the receiver, normal completion occurs when the NIC observes the last fragment of a message (marked explicitly). 
Even if earlier packets were lost, receiving the final one signals message completion and triggers a \textsf{CQE}.

When the final fragment never arrives, \name{} uses an application-specified timeout to avoid indefinite stalls.
Each \textsf{WQE} includes a timeout value that bounds how long it can remain active. 
If this deadline expires before complete data arrival, the NIC finalizes the \textsf{WQE} and generates a \textsf{CQE} indicating partial progress.

To track this, the NIC maintains a per-\textsf{WQE} byte counter that accumulates the payload size of successfully placed packets. 
This logic reuses existing DMA metadata and adds only minimal state. 
Upon timeout, the NIC reports this count to the application, allowing the upper layer (\eg, the collective engine) to proceed with partial data.

Timeouts are managed using per-\textsf{WQE} hardware timers, similar to those already implemented for retry or RNR timeout logic in reliable transports~\cite{mittal2018revisiting,wang2023srnic}.
These timers are reused but reinterpreted: instead of triggering retransmissions, they now bound execution time.

\paragraph{\em Early Completion via Preemption.}
\name{} introduces a form of early timeout via preemption.
If the receiver observes a packet from a newer message (with a higher \textsf{wqe\_seq}), it immediately finalizes the current message and begins processing the new one. 
This mechanism ensures timely progress and bounds per-\textsf{WQE} state, even when packets are delayed or reordered. 
Any subsequent packets from the older message are dropped, ensuring correctness.

\paragraph{\em Adaptive Timeout Estimation.}
Choosing a fixed timeout is challenging in distributed ML workloads, where network conditions and collective patterns vary widely~\cite{yang2022inflless, fu2024serverlessllm}.
To address this, \name{} includes an adaptive timeout mechanism that adjusts values over time.

After each collective operation, nodes record two key statistics: the elapsed time and the number of bytes successfully received, including both full and partial completions. 
These values are exchanged asynchronously across
the collective group and used to compute an empirical per-byte transfer cost (\eg, microseconds per kilobyte). 
Each node then proposes a timeout value for future iterations, derived by multiplying this cost by the message size.

Before the next invocation of the same collective on the same group, nodes aggregate the proposed values to form a group-wide timeout. 
They compute the median across all peers to reduce the impact of outliers (\eg, nodes experiencing transient loss or congestion). 
To further avoid oscillation, especially in small collectives, the group applies an exponentially weighted moving average (EWMA) to smooth the update: $T_{\text{new}} = \alpha \cdot T_{\text{median}} + (1 - \alpha) \cdot T_{\text{old}}$.
We use $\alpha = 0.2$, which balances responsiveness with stability. 
The resulting value becomes the canonical timeout estimate for future operations of the same collective and group.

If no historical observations are available---such as on the first invocation---\name{} initializes the timeout using the measured duration of a warmup collective executed during the bootstrap phase. Specifically, it sets: $T_{\text{initial}} = (1 + \gamma) \cdot T_{\text{warmup}} + \delta$, where $\gamma$ is a multiplicative safety margin (we use 0.25) and $\delta$ is a small additive slack (50\textmu s) to absorb short-term variance. This conservative baseline ensures that early iterations proceed reliably while timeout estimation converges.

Timeouts are applied at the granularity of individual RDMA operations. For collective algorithms with multiple phases, the total timeout budget is divided across phases: parallel steps share the same deadline, while sequential steps are assigned proportional slices, ensuring that the entire operation completes within the allotted time.

Finally, small control-plane messages---like handshakes and phase markers---are typically under one MTU and do not impact tail latency or bandwidth. 
\name{} routes them over the pre-existing reliable channel, avoiding unnecessary timeout logic and keeping the data path focused on large transfers.

\paragraph{\em Timeout Behavior Across Verbs.}
Timeout behavior follows the standard RDMA model: it applies only to the side that posts a \textsf{WQE}.

\begin{itemize}[leftmargin=*]
	\item \textsf{SEND}/\textsf{RECV} (two-sided verbs): Both sender and receiver post \textsf{WQEs} and each side attaches its own timeout.\vspace{-3pt}	
	\item \textsf{WRITE} (one-sided verb): Only the sender posts a \textsf{WQE} and sets a timeout. The receiver performs DMA but does not track time.\vspace{-3pt}
	\item \textsf{WRITE\_WITH\_IMM}: Behaves like a hybrid; both sides post \textsf{WQEs}, and timeouts are active on both ends.\vspace{-3pt}
	\item \textsf{READ}: The requester attaches a timeout. To avoid unnecessary transmissions, \name{} piggybacks this deadline in the request, allowing the responder to stop sending after the deadline.\vspace{-3pt}
\end{itemize}

\subsubsection{Congestion Control Semantics.~}
\label{sssec:cc}
In traditional RDMA transports, congestion control (CC) is tightly coupled with reliability: packet loss is treated as a congestion signal and triggers retransmission. 
\name{} eliminates retransmissions entirely, decoupling these two mechanisms. 
In this model, dropped packets no longer imply congestion, and feedback packets from the receiver (\eg, \textsf{ACKs} or \textsf{CNPs}) are interpreted purely as best-effort congestion signals, not as proofs of reliable delivery.

Despite this shift, \name{} remains fully compatible with the dominant RDMA CC schemes deployed in practice. 
ECN-based algorithms like DCQCN~\cite{dcqcn} rely on explicit switch marks and \textsf{CNPs} generated for packets that do arrive; their control loops operate unchanged. 
Delay-based controllers such as TIMELY~\cite{mittal2015timely} and Swift~\cite{kumar2020swift} compute RTT from timestamped feedback packets, which \name{} continues to generate for received packets; lost packets yield no feedback. 
Likewise, telemetry- and credit-based schemes such as HPCC~\cite{li2019hpcc} and EQDS~\cite{olteanu2022edge} depend on in-band telemetry or explicit credit messages---none of which require reliable delivery of every data packet.

\subsection{Lightweight Data Recovery \& Loss Mitigation}
\label{ssec:recovery-mitigation}

Since \name{} operates without retransmissions or in-order delivery, some data loss is expected. 
Rather than masking this loss via heavyweight transport-layer repair, \name{} leverages the inherent robustness of ML workloads and introduces a lightweight software mechanism to mitigate its impact. 
The key goal is to prevent localized packet drops from introducing correlated corruption in model state or gradient tensors.

\paragraph{\em Loss Amplification From Spatial Clustering.}
In a naive design, each packet carries a contiguous slice of a tensor. 
If this packet is lost, its entire span of values is zeroed during placement. 
This spatially clustered loss disproportionately affects model quality: adjacent values often correspond to neighboring neurons or channels, and the resulting distortion can destabilize training or degrade inference accuracy. 
Prior work has shown that structured randomness---such as that introduced by the Hadamard Transform~\cite{pratt1969hadamard}---can spread such errors across a tensor and preserve convergence even under significant loss~\cite{warraich2025optireduce,wang2024towards,li2024thc}.

\begin{tcolorbox}[
    colback=blue!5!white,
    colframe=black,
    title={},
    fonttitle=\bfseries,
    fontupper=\footnotesize, 
    sharp corners,
    boxrule=0.4mm,
    coltitle=black,
    enhanced jigsaw,
    drop shadow={black!50!white}
]
\textbf{
INFO: Hadamard Transform~\cite{pratt1969hadamard}.} {It is an orthogonal linear mixing operation that disperses each input element across all output coefficients. This spreads local errors uniformly and preserves tensor norms, making it effective for mitigating sparse loss.}
\end{tcolorbox}

\paragraph{\em (a) Block-Wise Encoding for Compute Efficiency.}
To reduce computational overhead, \name{} applies the Hadamard Transform in a block-wise manner. 
Each tensor is logically divided into $B$ blocks of $p$ elements (typically matching the per-packet MTU size), and each block is transformed independently. 
Because the transform is linear, encoded tensors can be aggregated or reduced without decoding---a useful property for collectives (like AllReduce). 
Block-wise encoding significantly lowers GPU compute cost (\S\ref{ssec:microbenchmarks}) but is not sufficient by itself: if a packet carries an entire encoded block and is lost, all $p$ coefficients from that block are erased, nullifying the transform's resilience benefit.

\paragraph{\em (b) Stride-based Packet Interleaving to Improve Recovery.}
To prevent this failure mode, \name{} introduces a stride-based layout that interleaves encoded blocks across packets. 
After transforming each block, packets are constructed by selecting a fraction of coefficients from multiple blocks, rather than all coefficients from a single block. 
Specifically, a stride parameter $S$ determines how many blocks contribute to a packet: each packet carries $p/S$ elements from each of $S$ blocks, for a total of $p$ elements. 
This interleaving spreads the impact of a lost packet across many blocks, reducing the distortion experienced by any single one.

This layout is efficiently implemented via Scatter–Gather Entries (SGEs), an RDMA feature~\cite{nvidia-rdma-manual} that allows non-contiguous memory regions to be transmitted in a single message. 
When striding is enabled, each packet header includes $S$ and a per-packet offset,
allowing the receiver to perform correct placement without coordination or ordering guarantees---extending the self-describing packet abstraction (\S\ref{sssec:data-delivery}).

With maximal striding ($S = p$), each packet contains one coefficient from each of $p$ blocks. Losing a packet in this regime zeroes one element per block, converting clustered loss into sparse noise. 
The inverse Hadamard Transform uniformly distributes this residual error across each affected block. 
The stride parameter $S$ thus allows \name{} to trade off dispersion strength against complexity: higher $S$ improves data recovery, but mixes more block elements per packet. 
We evaluate the benefits of this design in \S\ref{ssec:microbenchmarks}, showing that it preserves accuracy even under nontrivial packet loss.

\subsection{Realizing \name{} on Existing RDMA NICs}
\label{ssec:mapping-nics}
Although \name{} defines new transport semantics, many of its mechanisms align with capabilities already present in modern RDMA NICs~\cite{nvidia-rdma-manual}.
Realizing \name{} therefore requires only modest deltas: removing features that are no longer needed, reusing existing abstractions for new purposes, and introducing small software or header-level extensions. 
We describe how \name{} maps onto two representative platforms---SRNIC and commodity RoCE NICs---while highlighting where these NICs fall short and how \name{} bridges those gaps.

\paragraph{\em SRNIC: Removing Reliability, Reusing Metadata and Timers.}
SRNIC~\cite{wang2023srnic} already contains several building blocks that \name{} relies on---per-packet metadata for direct placement (remote address, offset, sequence ID), and an in-place DMA pipeline that bypasses reassembly. 
These features allow \name{}'s self-describing delivery semantics to be supported without structural changes to the datapath. 
Likewise, SRNIC's per-\textsf{WQE} timers, originally intended for retransmissions and RNR backoff, are repurposed to enforce \name{}'s bounded completion model with byte tracking.

However, SRNIC's original design assumes reliable delivery. 
Implementing \name{} requires removing the full reliability subsystem---including bitmap tracking, outstanding-request tables, and loss-recovery state machines---which no longer serve a purpose and impede tail performance. 
\name{} thus simplifies the NIC by eliminating these components entirely. 
The only new NIC-visible field is a 2-byte stride parameter to support recovery (\S\ref{ssec:recovery-mitigation}), while congestion control continues to use the existing control queue (\textsf{CtrlQ}) to surface \textsf{ECN} and pacing signals to software. 
The result is a smaller, faster, and more resilient datapath that preserves compatibility with the RDMA programming model.

\paragraph{\em RoCE w/ UC: Software Approximation of \name{} on Fixed-Function Hardware.}
Commodity ConnectX-class RoCE NICs offer no datapath programmability, and their UC transport enforces in-order delivery for multi-packet messages---a semantics incompatible with \name{}'s out-of-order, best-effort model. Realizing \name{} therefore requires a software approximation that works within these hardware constraints.

The key approximation is forcing every fragment to be a single-packet \textsf{WRITE}. 
Each MTU-sized block is issued as an independent \textsf{WRITE\_WITH\_IMM} carrying explicit placement metadata, ensuring that the NIC never triggers its built-in ordering or reassembly logic. 
This preserves \name{}'s semantics despite UC's fixed in-order behavior.

Completion and timeout semantics are reimplemented entirely in software, using immediate values to identify fragments and timer queues to enforce deadlines.
 To prevent corruption after timeout---something commodity NICs cannot guard against---we rely on Memory Windows (MWs) with per-operation rkeys, allowing receivers to revoke write access mid-collective and block late \textsf{WRITEs}.
Congestion control is likewise implemented in software. 
Fragment-level feedback packets guide pacing decisions in lieu of NIC-managed rate control~\cite{zhou2025extensible}; lost fragments yield no feedback, which integrates cleanly with \name{}'s best-effort model.

Although this realization incurs modest CPU overhead, it preserves all core \name{} semantics---independent placement, bounded completion, and explicit pacing---without requiring firmware or driver changes. 
This makes \name{} immediately deployable on existing RoCE networks while highlighting the limitations of today's NICs and the architectural simplifications \name{} enables.

%% file: sections/implementation.tex
\section{Implementation}
\label{sec:implementation}

We implement \name{} on two platforms: a software prototype running over commodity RoCE NICs to evaluate end-to-end training and inference, and a hardware prototype on an FPGA-based SmartNIC to assess area, state overhead, and transport-level scalability.

\paragraph{\em RoCE Software Prototype.}
We implement \name{} as a new transport backend in NVIDIA's NCCL (\textsf{v2.23.4–1}) via the Net plugin interface~\cite{jeaugey2017nccl}, enabling out-of-the-box compatibility with DeepSpeed (\textsf{v0.18.2}), PyTorch,
and vLLM (\textsf{v0.9.1}).
The integration adds fewer than 500 lines to NCCL, with the transport logic written in about 8K lines of C++.

All control-plane logic—including completion tracking, timeout management, and congestion control—is implemented in software. A dedicated timer thread manages deadlines, while congestion control uses EQDS~\cite{olteanu2022edge}, with the sender pacing transmissions based on per-fragment ACKs.

Hadamard transforms for error recovery are implemented on GPU using an optimized CUDA kernel from HazyResearch~\cite{structured-nets}, applied block-wise during encoding/decoding.

\paragraph{\em FPGA Hardware Prototype.}
The hardware implementation of \name{} is built on the AMD Alveo U250 FPGA using Coyote-v2, an open-source RoCEv2-compatible SmartNIC shell~\cite{ramhorst2025coyote}.
We synthesize the design in Vivado \textsf{2022.1} and target 10K QPs to match common transport scalability needs.

We evaluate \name{}'s hardware resource usage by removing Coyote's built-in reliability mechanisms---retransmission logic, outstanding-request tables, bitmaps, and reorder buffers---and adding minimal per-\textsf{WQE} state for timeouts and byte tracking. 
The transport pipeline reuses Coyote's existing support for self-describing placement, and we extend the packet header by 2 bytes to support stride placement for recovery.

For comparison, we also synthesize three baselines: (a) IRN/Falcon, which uses a \SI{1.2}{MB} reorder buffer and reconstructed QP state based on prior work. (b) SRNIC, using QP metadata and extensions as described in the original paper. (c) UCCL, which requires no hardware changes and runs atop base RoCE.
This setup allows us to directly compare datapath area, path delay, and QP state across all designs in \S\ref{ssec:microbenchmarks}.

%% file: sections/evaluations.tex
\section{Evaluating \name{}}
\label{sec:evaluation}

We evaluate \name{} to validate our central claim: simplifying RDMA transport for ML workloads improves tail latency, reduces hardware overhead, and enhances system resilience. 
Our evaluation covers three dimensions—latency, efficiency, and fault tolerance—using both microbenchmarks and end-to-end distributed workloads on a real-world Cloud cluster and an FPGA-based prototype.

\begin{figure*}[t]
    \centering
    \includegraphics[width=\textwidth]{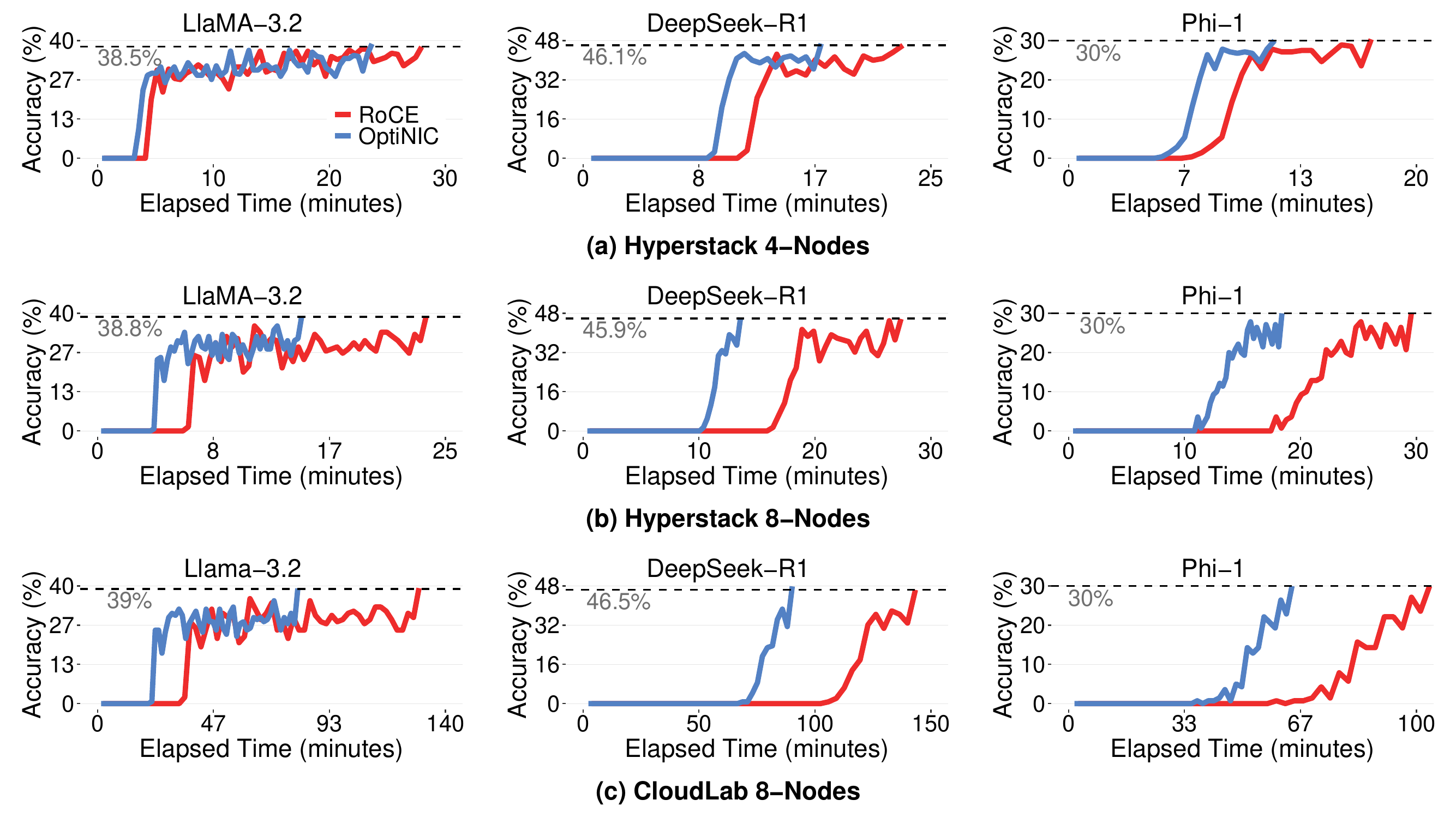}
    \vspace{-20pt}
    \caption{End-to-end convergence time-to-accuracy of RoCE and \name{} for different models and cluster environments.}
    \vspace{-15pt}
    \label{fig:training-tta}
\end{figure*}

\begin{figure*}[t]
    \centering
    \includegraphics[width=\textwidth]{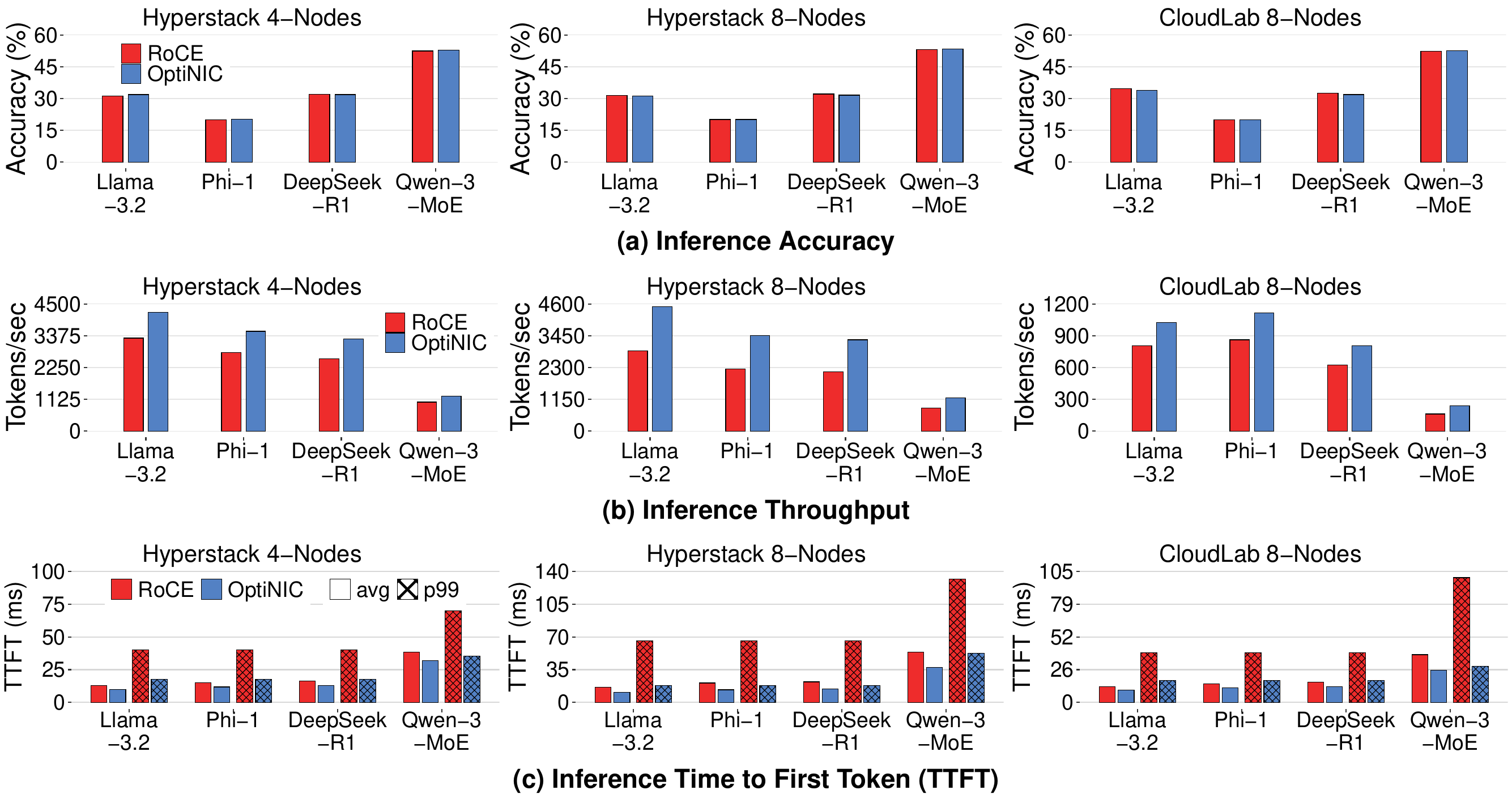}
    \vspace{-22pt}
    \caption{Inference accuracy, throughput, and time-to-first-token (TTFT) across models and cluster environments.}
    \vspace{-15pt}
    \label{fig:inference-unified}
\end{figure*}

\subsection{Experimental Setup}

\subsubsection{Test Environments.~} 

Our experiments are conducted on three environments from an academic cloud, CloudLab~\cite{cloudlab}, and a commercial cloud vendor, Hyperstack~\cite{hyperstack}.
On CloudLab, we provision an 8-node \textsf{r7525}~\cite{cloudlab-r7525} cluster, with each machine featuring dual AMD EPYC~7542 CPUs (64~cores, 2.9\,GHz), 512\,GB DDR4 ECC memory, with an NVIDIA Tesla~V100S GPU (32\,GB). 
Networking is provided by dual-port Mellanox ConnectX-5 NICs (PCIe~Gen4), connected via a 25\,Gbps Ethernet fabric.
All nodes run Ubuntu~22.04 and NCCL~\textsf{v2.23.4--1} with default configurations and \name{} with NCCL on the SmartNICs (\S\ref{sec:implementation}).
To emulate realistic multi-tenant conditions, we introduce controlled background traffic that reflects RDMA network behavior reported in prior works~\cite{guo2016rdma,mittal2018revisiting}.
On Hyperstack, we provision 4 and 8 node \textsf{H100-80G-PCIe} clusters with each machine featuring 28 CPU cores, 180 GB DRAM, and an Nvidia H100 GPU with 80\,GB HBM3 memory and PCIe Gen5 interconnect.

\subsubsection{Baselines, Workloads, and Cluster Setup.~} 
We evaluate three recent open-source LLMs---Llama-3.2-1B~\cite{grattafiori2024llama}, Phi-1-1B~\cite{gunasekar2023textbooks}, and DeepSeek-R1-Distill-Qwen-1.5B~\cite{guo2025deepseek}---chosen to span different architectures and model families. 
For training, we fine-tune each model on the ARC-Challenge dataset~\cite{clark2018think} using DeepSpeed with ZeRO-3 parallelism, keeping all other training hyperparameters fixed across transports. 
For inference, we again use ARC-Challenge prompts and measure end-to-end generation throughput (tokens/sec), latency (TTFT), and accuracy using vLLM with all three models served using Tensor + Pipeline parallelism. 
For inference, we also serve the Qwen-3-30B MoE model~\cite{yang2025qwen3} using Tensor+Expert parallelism to exercise MoE-specific communication patterns.

Our primary baseline is RoCEv2 with RC QPs, which enforces retransmissions, in-order delivery, and strict completion semantics at the NIC level—representative of production datacenter RDMA deployments. 
We also benchmark \name{}'s performance against IRN~\cite{mittal2018revisiting}, SRNIC~\cite{wang2023srnic}, Falcon~\cite{singhvi2025falcon}, UEC~\cite{uec2025spec}, and UCCL~\cite{zhou2025extensible}. 
Open-source implementations of these are unavailable or incompatible with cloud environments and are therefore excluded from end-to-end performance runs.\footnote{We attempted to include UCCL~\cite{zhou2025extensible} in our end‑to‑end evaluation, but encountered a bug; the issue has been confirmed by the developers and is currently being fixed.}
To assess the resilience of these implementations at scale, we evaluate their soft-error susceptibility using the Xilinx SEU Estimator \textsf{v2023.1}~\cite{amd-seu-estimator}. 
Following datacenter deployment guidelines, we model a 15,000-node cluster operating at a junction temperature of \SI{100}{\celsius}
~\cite{smartnic-power-profile-1,smartnic-power-profile-2}, enabling a comparative analysis of reliability across all transport designs under realistic large-scale conditions.

\subsection{End-to-End Performance}
\label{ssec:e2e-evals}

\subsubsection{Distributed Training.~}
\Cref{fig:training-tta} reports convergence trajectories for Llama-3.2-1B, Phi-1-1B, and DeepSeek-R1-Distill-Qwen-1.5B when fine-tuned on ARC-Challenge with ZeRO-3 parallelism. 
Across all three cloud environments and models, \name{} reduces TTA by $1.6\times$ on average relative to RoCE. Larger configurations benefit more: the 8-node setups yield up to $2\times$ improvement, particularly on Hyperstack where H100 GPUs shift the bottleneck toward communication. While CloudLab shows larger raw communication gains, its V100 GPUs limit end-to-end speedups. 
These gains arise because RoCE's Go-Back-N loss recovery briefly halts progress for all nodes, even when only a single packet is dropped.
In contrast, \name{} continues making forward progress within each bounded window, avoiding these global pauses. 
Final accuracy is unchanged across models, and in some cases even slightly improves. 
For example, DeepSeek-R1 achieves around a $1.2\%$ higher final accuracy with \name{}, as the small, bounded perturbations introduced by random and infrequent packet drops act as a mild form of regularization---similar in spirit to noise injection or dropout---and can occasionally enhance generalization.

\subsubsection{Distributed Inference.~}
\Cref{fig:inference-unified} shows inference accuracy, throughput (tokens/sec), and latency (TTFT) across all evaluated models and cloud environments.
Interestingly, the Qwen-3-30B MoE model also shows a small accuracy increase with \name{}. 
In MoE inference, small activation-level perturbations can change which experts are selected, occasionally producing outputs that score slightly higher.
As vLLM serving is far less communication-intensive than ZeRO-3 training, the gains of \name{} are correspondingly more moderate but consistently significant.
\Cref{fig:inference-unified}a shows that \name{} inference accuracy remains effectively unchanged across the board (differences $<0.2\%$) against tail-inducing reliable RoCE transport across all models and environments.
In \Cref{fig:inference-unified}b, across the non-MoE models, \name{} improves throughput by roughly $28$–$60\%$ over RoCE. 
Finally, average TTFT improves slightly, whereas tail (p99) latency drops sharply across all models (2--3.5$\times$) as shown in \Cref{fig:inference-unified}c, consistent with \name{}’s tail-optimal design.
Notably, the largest gains in both throughput and TTFT tail appear on the Hyperstack 8-node configuration, where the stronger H100 GPUs make communication the dominant bottleneck and \name{}'s benefits manifest most prominently.

\input{sections/microbenchmarks}

%% file: sections/microbenchmarks.tex
\subsection{Microbenchmarks}
\label{ssec:microbenchmarks}

\subsubsection{\name{} is up to 2.5$\times$ faster than RoCE across all collectives and message sizes.~}
\begin{figure}[t]
    \centering
    \includegraphics[width=\linewidth]{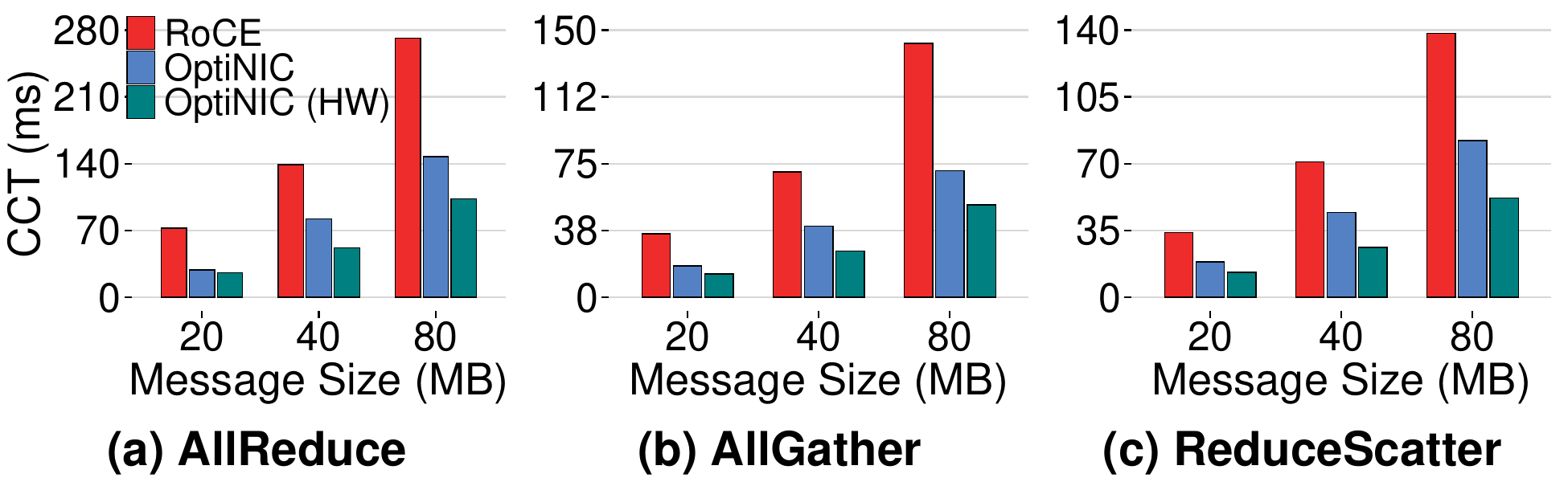}
    \vspace{-20pt}
    \caption{Collective communication time: comparison across transports, message sizes, and collective types. (Stddev, RoCE:$\pm$35, \name{}:$\pm$12, and \name{} (HW):$\pm$3.} 
    \label{fig:message-size}
    \vspace{-10pt}
\end{figure}

\Cref{fig:message-size} compares \name{} and RoCE across tensor sizes from 20--80~MB for AllReduce, AllGather, and ReduceScatter collectives. 
RoCE’s latency grows steeply with size, reflecting the cumulative cost of retransmissions and completion dependencies. 
In contrast, \name{} scales smoothly: latency increases only moderately, remaining mostly linear, and consistently delivering a 1.6--2.5$\times$ speedup over RoCE across the tested sizes and collectives. 
To emulate a hardware deployment of \name{} (HW), we subtract software overheads (segmentation, timers, pacing) from the RoCE-based prototype, isolating the transport's performance contribution.
Despite omitting retransmissions, observed loss stays under 1\% on average for the tensors. 
Under a more aggressive timeout for large tensors—accepting 4--5\% loss—\name{} achieves up to 5$\times$ lower latency than RoCE (not shown), highlighting the performance headroom unlocked when applications tolerate slightly higher loss. 

\begin{figure}
\centering
\includegraphics[width=1.0\columnwidth]{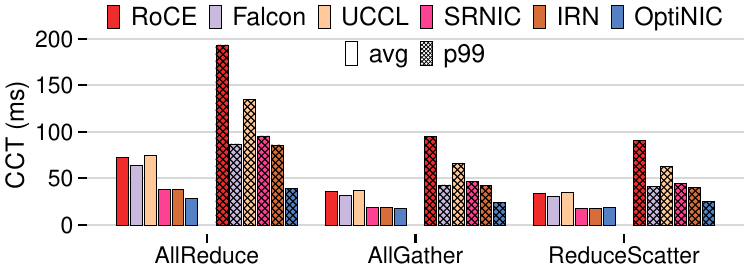}
\vspace{-20pt}
\caption{Collective completion time: average and tail comparison for different transports.}
\label{fig:transport-collective-fct-speedup}
\vspace{-18pt}
\end{figure}

\subsubsection{\name{} delivers the fastest collective completion times across all transports.~}

As shown in~\Cref{fig:transport-collective-fct-speedup}, \name{} achieves the lowest collective communication times across all three collectives---AllReduce, AllGather, and ReduceScatter---delivering both the smallest average Collective Completion Time (CCT) and the lowest tail latency (p99) among all transports. 
In contrast, RoCE, Falcon, and UCCL exhibit similar mean performance, but their tail latencies remain significantly higher: Falcon and UCCL match RoCE on average yet their tail rises to levels comparable to IRN and SRNIC, highlighting persistent head-of-line blocking and retry overheads. 
IRN and SRNIC modestly reduce mean CCT but still suffer from large p99 spikes, particularly for AllReduce, where tail latencies exceed 100--150\,ms. 
By eliminating retransmissions and reordering entirely, \name{} avoids these tail-amplifying effects and consistently delivers both fast and tightly bounded completion times.

\begin{figure}
\centering
\includegraphics[width=1.0\columnwidth]{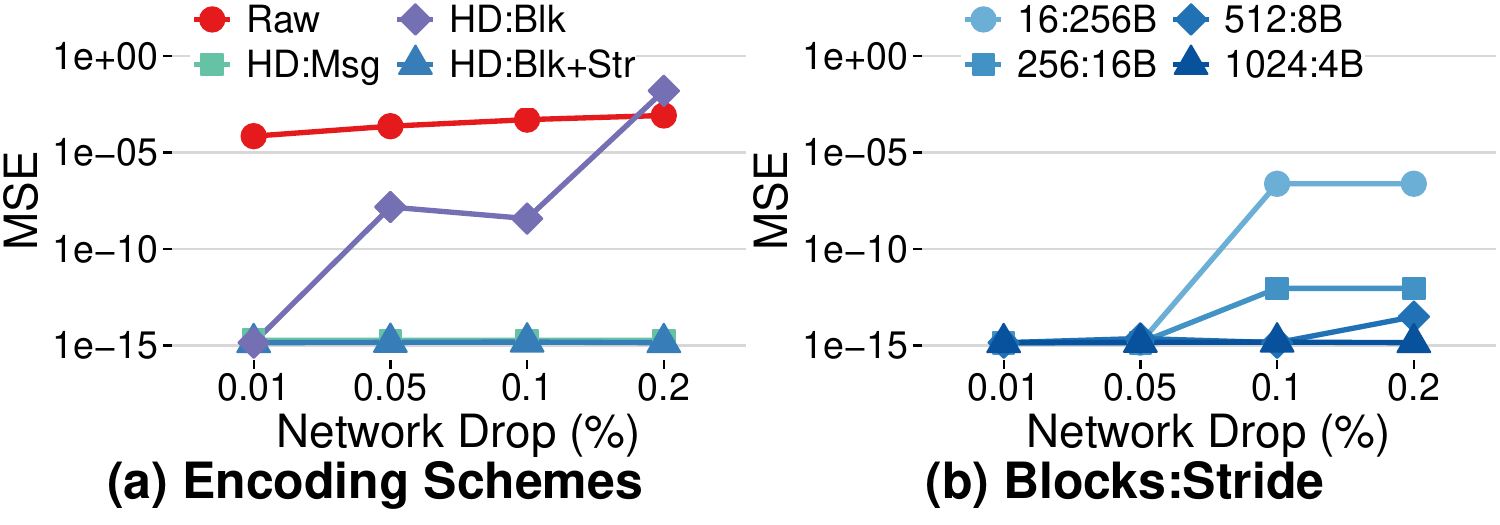}
\vspace{-20pt}
\caption{Comparison of MSE with Hadamard on (a) different configurations and (b) stride parameters.}
\label{fig:loss-recovery}
\vspace{-5pt}
\end{figure}

\begin{table}[t]
\centering
\small
\resizebox{0.999\columnwidth}{!}{
\begin{tabular}{lcccc}
\toprule
\multicolumn{1}{r}{\bf \#Splits $\rightarrow$} & {\bf 1} & {\bf 4} & {\bf 16} & {\bf 64} \\
\midrule
{\bf Time (ms)} 
    & \textcolor{red}{$22.1 \pm 0.08$}
    & $19.9 \pm 0.04$
    & $17.5 \pm 0.07$
    & $8.4 \pm 0.13$ \\
\bottomrule
\end{tabular}}
\vspace{-10pt}
\caption{\bf Mean $\pm$ std runtime of Hadamard across different split counts for a 128\,MB message.}
\vspace{-18pt}
\label{tab:hadamard-splits}
\end{table}

\subsubsection{Hadamard with stride delivers robust, efficient loss dispersion.~}

\Cref{tab:hadamard-splits} shows that a Hadamard transform on raw 128\,MB message is the most expensive configuration, while splitting the tensor into 64 blocks reduces runtime by $2.5\times$, motivating block-level processing. 
\Cref{fig:loss-recovery}a examines the resulting resilience tradeoffs: Raw (no coding) and full-message Hadamard (\textsf{HD:Msg}) behave as expected, with the latter achieving near-ideal MSE at the highest cost. 
Block-wise Hadamard (\textsf{HD:Blk}) is far cheaper but can catastrophically amplify error because a lost packet removes all encoded coefficients for a block, making recovery impossible. 
Adding striding (\textsf{HD:Blk+Str}) disperses coefficients across packets so each loss removes only one position per block, producing MSE comparable to full-message Hadamard at much lower overhead. 
\Cref{fig:loss-recovery}b also shows that resilience improves with increasing stride: small strides couple many coefficients per packet and increase MSE, whereas maximal dispersion yields near–ideal reconstruction across all drop rates. 
Overall, \name{}’s \textsf{HD:Blk+Str} design matches the robustness of full-message transforms at a fraction of the computational cost, making striding essential for resilient block-wise encoding.

\begin{table}[t]
\centering
\small
\resizebox{0.999\columnwidth}{!}{
\begin{tabular}{@{}l|r|r|r|r|r|>{\columncolor[HTML]{cfe2f3}}r}
\toprule
\textbf{Metric}
    & \multicolumn{1}{l|}{\textbf{RoCE}}
    & \multicolumn{1}{l|}{\textbf{IRN}}
    & \multicolumn{1}{l|}{\textbf{SRNIC}}
    & \multicolumn{1}{l|}{\textbf{Falcon}}
    & \multicolumn{1}{l|}{\textbf{UCCL}}
    & \multicolumn{1}{>{\columncolor[HTML]{cfe2f3}}l}{\textbf{\name{}}} \\
\midrule

NIC State per QP
    & \SI{407}{B} 
    & \SI{596}{B}  
    & \SI{242}{B}  
    & \SI{350}{B}  
    & \SI{407}{B}  
    & \textbf{\SI{52}{B}} \\

Max. QPs        
    & 10K 
    & 8K 
    & 20K 
    & 12K 
    & 10K 
    & \textbf{80K} \\

Cluster Size      
    & 5K 
    & 4K 
    & 10K 
    & 6K 
    & 256 
    & \textbf{40K} \\
\bottomrule
\end{tabular}
}
\vspace{-8pt}
\caption{Comparison of transport protocols across NIC state, QP count, and cluster scalability.}
\vspace{-5pt}
\label{tab:qp-scalability}
\end{table}

\subsubsection{\name{} achieves order-of-magnitude higher QP scalability with minimal QP state.~}
\name{} delivers the highest scalability among all evaluated transports, as shown in \Cref{tab:qp-scalability}, by reducing per-QP NIC state to just 52\,B---an order of magnitude smaller than RoCE (407\,B), IRN (596\,B), SRNIC (242\,B), Falcon (350\,B), and UCCL (407\,B). 
This dramatic reduction enables \name{} to support up to 80K active QPs within the same SRAM budget (\SI{4}{MB}) that limits existing designs to between 8K and 20K QPs. 
As a result, \name{} scales collective training to 40K nodes, far exceeding the limits of RoCE (5K), IRN (4K), and Falcon (6K), and doubling SRNIC’s 10K-node scale. UCCL scales even more poorly, as it opens 256 connections per peer—compared to the default 2 for all other schemes—which quickly exhausts NIC resources at large cluster sizes.

\begin{table}[t]
\centering
\small
\resizebox{0.999\columnwidth}{!}{
\begin{tabular}{@{}l|r|r|r|r|r|>{\columncolor[HTML]{cfe2f3}}r}
\toprule
\textbf{Metric} 
    & \multicolumn{1}{l|}{\textbf{RoCE}}
    & \multicolumn{1}{l|}{\textbf{IRN}}
    & \multicolumn{1}{l|}{\textbf{SRNIC}}
    & \multicolumn{1}{l|}{\textbf{Falcon}}
    & \multicolumn{1}{l|}{\textbf{UCCL}}
    & \multicolumn{1}{>{\columncolor[HTML]{cfe2f3}}l}{\textbf{\name{}}} \\
\midrule

LUT & 312.4K & 319.6K & 304.5K & 309.8K & 312.4K & \textbf{298.4K} \\
LUTRAM & 23.3K  & 24.2K  & 22.5K  & 23.1K  & 23.3K  & \textbf{21.7K} \\
FF & 562.1K & 573.1K & 551.5K & 559.2K & 562.1K & \textbf{543.0K} \\
\midrule

BRAM & 1.5K   & 2.2K   & 0.9K   & 1.6K   & 1.5K   & \textbf{0.5K} \\
Power (W) & 34.7   & 35.9   & 33.5   & 34.3   & 34.7   & \textbf{32.5} \\
\midrule

MTBF (hrs) & 42.8   & 30.9   & 57.8   & 40.5   & 42.8   & \textbf{80.5} \\
\bottomrule
\end{tabular}
}
\vspace{-8pt}
\caption{
Comparison of hardware resource utilization and resilience (MTBF) across RDMA NIC architectures.
}
\vspace{-14pt}
\label{tab:hardware-summary}
\end{table}

\subsubsection{\name{} delivers maximum hardware resilience with the smallest NIC footprint.~}

\Cref{tab:hardware-summary} shows that \name{} achieves the smallest hardware footprint and highest resilience among all evaluated NIC transports. 
Compared to RoCE, IRN, SRNIC, Falcon, and UCCL under a 10K-QP configuration synthesized for AMD Alveo U250, \name{} reduces LUT usage to 298.4K, LUTRAMs to 21.7K, and FFs to 543.0K---representing up to 6.6\%, 10.2\%, and 5.2\% savings, respectively. 
Most notably, \name{} cuts BRAM consumption to just 0.5K (a 63--73\% reduction versus RoCE and IRN) and lowers power draw to 32.5\,W. 
By eliminating retransmission, reordering, and per-QP window state, \name{} maintains only 52\,B of per-QP context and thus achieves the highest MTBF of 80.5 hours---nearly 2$\times$ better than RoCE and IRN---demonstrating that simplifying transport-layer hardware simultaneously improves efficiency and robustness.

%% file: sections/limitations.tex
\section{Discussion and Future Work}
\label{sec:discussion}

\paragraph{\em Deployment on DPUs and SmartNICs.}
While our RoCE prototype executes its software control plane on the host CPU, nothing in \name{}'s design requires host participation. The timeout manager, per-fragment ACK processing, and software congestion controller all operate on simple, event-driven logic that can run unmodified on modern DPUs and SmartNICs, which already expose programmable ARM clusters or P4/XDP execution units. Offloading these components moves \name{}’s control path closer to the NIC datapath, reducing host overhead and enabling tighter pacing loops without altering any transport semantics.

\vspace{-2pt}
\paragraph{\em Reproducibility and Sources of Nondeterminism.}
A known limitation of best-effort transport for distributed training and inference is reduced reproducibility: transient packet losses may lead to nondeterministic training (or inference) dynamics across runs. However, this trade-off is increasingly acceptable in large-scale LLM workloads, where nondeterminism is already prevalent due to numerical instabilities, asynchronous kernel execution, and dynamic parallelism. Even models configured for determinism (\eg, fixed random seeds and zero sampling) often exhibit nontrivial variance in convergence accuracy and output behavior~\cite{atil2024non}. In practice, this means that strict reproducibility is rarely achieved end-to-end and system-induced variance is one of many contributing factors. Nevertheless, \name can optionally log missing offsets or byte ranges per step, enabling post hoc debugging or reproduction of loss patterns when needed.

\vspace{-2pt}
\paragraph{\em Beyond Distributed Training: Broader Applicability.}
While our work focuses on distributed training and inference, the same principles can benefit a broader class of applications that value timeliness over strict reliability.  
Latency-critical and soft real-time systems such as online recommendation services, interactive analytics, and real-time media streaming (\eg, video conferencing) often tolerate minor data loss or approximation in exchange for bounded response times.  
Applying bounded-loss transport semantics to these domains opens an exciting direction for rethinking communication not as a strictly reliable service, but as a controllable dimension in system design, where performance, efficiency, and accuracy can be balanced according to application goals.
\name{} demonstrates that reliability is not a universal requirement, but a workload-dependent choice. Relaxing the traditional transport guarantees can yield tangible benefits in tail latency and scalability. We believe this opens up a broader space of domain-specific transport designs that trade reliability for performance, tuned to the needs of the workloads.

%% file: sections/related-work.tex
\section{Related Work}
\label{sec:related}

\paragraph{\em RDMA Transports and NIC Architecture.}
IRN~\cite{mittal2018revisiting} removes PFC by introducing selective-repeat reliability in the NIC, using bitmap tracking and SACK-based retransmissions to tolerate loss in RoCE clusters.  
SRNIC~\cite{wang2023srnic} simplifies the NIC datapath by eliminating WQE caching and shifting retransmissions and reordering to host software, improving QP density and reducing NIC memory pressure.  
UCCL~\cite{zhou2025extensible} continues this trend by offloading transport control---congestion control, flow scheduling, and multipath routing---into software, treating the NIC as a streamlined datapath.  
Conversely, Falcon~\cite{singhvi2025falcon} embraces NIC complexity with fast retransmissions, delay-based congestion control, and hardware multipath routing, while UEC~\cite{uec2025spec,hoefler2025ultra} proposes a clean-slate transport for AI workloads with packet spraying and hybrid congestion control.  
Despite their differences, these designs preserve strict reliability semantics and recover every lost packet.  
In contrast, \name{} removes packet-level recovery entirely and introduces bounded-loss completion semantics in software, yielding a best-effort RDMA transport for ML workloads.

\vspace{-2pt}
\paragraph{\em Collective Communication and ML Systems.}
NCCL~\cite{jeaugey2017nccl}, MSCCL~\cite{msccl}, and UCC~\cite{venkata2024unified} accelerate collective operations through topology-aware scheduling, fused kernels, and unified CPU/GPU/DPU interfaces.  
SHArP~\cite{graham2016scalable} performs hierarchical in-network reductions, while SwitchML~\cite{sapio2021scaling} and OmniReduce~\cite{fei2021efficient} offload aggregation into programmable switches or network dataplanes.  
Recent systems such as MSCCL++~\cite{shah2025msccl++}, NCCLX~\cite{si2025collective}, and MCCS~\cite{wu2024mccs} further improve collective scheduling and overlap on large GPU clusters.  
These approaches optimize or offload collective algorithms but fundamentally assume a reliable, in-order transport.  
\name{} differs by rethinking the transport itself: it provides a best-effort RDMA system that supports all collective patterns without specialized switches or full packet reliability.

\vspace{-2pt}
\paragraph{\em Lossy Transports and Approximate Communication for ML.}
Approximate communication techniques like Top-$k$ sparsification~\cite{stich2018sparsified} and quantization methods such as QSGD~\cite{alistarh2017qsgd} and TernGrad~\cite{wen2017terngrad} reduce gradient traffic with error compensation.
THC~\cite{li2024thc} enables homomorphic aggregation over quantized updates, while MLT~\cite{wang2024towards} explores bounded-loss behavior tuned for ML workloads.
OptiReduce~\cite{warraich2025optireduce} mitigates tail effects in AllReduce via time-bounded execution with adaptive recovery.
These approaches focus on software-level techniques and primarily target AllReduce.
In contrast, \name{} provides hardware-level loss tolerance in the RDMA datapath, generalizes to all collectives and parallelisms, and supports both training and inference while saturating modern 100--400,Gbps links.

%% file: sections/conclusion.tex
\section{Conclusion}
\label{sec:conclusion}

As ML systems scale to thousands of GPUs and flows, the assumption that transports must guarantee perfect delivery becomes increasingly misaligned with workload needs. 
Beyond performance, large NIC state and retransmission logic also reduce resilience and increase fault risk. 
\name{} shows that correctness in distributed ML does not require full reliability, but rather timely, bounded progress. 
By removing retransmissions and in-order delivery, and introducing adaptive, timeout-driven completion, \name{} simplifies the NIC while allowing collectives to proceed without waiting for stragglers. 
This design yields 1.8--2.5$\times$ faster collectives, 2.7$\times$ lower BRAM usage, and nearly 2$\times$ higher fault resilience. 
These benefits extend to end-to-end workloads: \name{} halves ZeRO-3 time-to-accuracy, improves inference throughput by 1.6$\times$, and reduces tail TTFT by 3.5$\times$, all without affecting accuracy. By prioritizing time-bounded progress over strict reliability, \name{} enables a faster, more scalable transport layer for ML.